\documentclass[aps,prl,twocolumn,superscriptaddress,showpacs]{revtex4}
\usepackage{hyperref}
\usepackage{amsmath}
\usepackage{amsfonts}
\usepackage{amssymb}
\usepackage{graphicx}
\begin{document}
\title{Effect of surface defects and few-atomic steps on the local density of states of the atomically-clean surface of topological insulator Bi$_2$Se$_3$}
\author{A.Yu. Dmitriev}
\email[]{dmitrmipt@gmail.com}
\affiliation{Kotel'nikov IRE RAS, Mokhovaya 11, bld.7, 125009 Moscow, Russia}
\affiliation{Moscow Institute of Physics and Technology, 141700 Dolgoprudny, Russia}

\author{N.I. Fedotov}
\affiliation{Kotel'nikov IRE RAS, Mokhovaya 11, bld.7, 125009 Moscow, Russia}
\affiliation{Moscow Institute of Physics and Technology, 141700 Dolgoprudny, Russia}

\author{V.F. Nasretdinova}
\affiliation{Kotel'nikov IRE RAS, Mokhovaya 11, bld.7, 125009 Moscow, Russia}

\author{S.V. Zaitsev-Zotov}
\affiliation{Kotel'nikov IRE RAS, Mokhovaya 11, bld.7, 125009 Moscow, Russia}
\affiliation{Moscow Institute of Physics and Technology, 141700 Dolgoprudny, Russia}

\date{\today}
\pacs{73.25.+i, 73.20.At, 71.30.+h, 74.55.+v}

\begin{abstract}
The results of ultra-high vacuum low-temperature scanning-tunneling microscopy (STM) and spectroscopy (STS) of atomically clean (111) surface of the topological insulator Bi$_2$Se$_3$ are presented. We observed several types of new subsurface defects whose location and charge correspond to p-type conduction of grown crystals. The sign of the thermoelectric effect also indicates p-type conduction. STM and STS measurements demonstrate that the chemical potential is always located inside the bulk band gap. 
We also observed changes in the local density of states in the vicinity of the quintuple layer steps at the studied surface. This changes correspond either to the shift of the Dirac cone position or to the shift of the chemical potential near the step edge.
\end{abstract}
\maketitle


Topological insulators (TIs) are the new class of materials that has many intriguing physical properties. Being insulators in the bulk, these
 materials always have gapless edge (1D) or surface (3D) states which are protected by the time-reversal symmetry \cite{HasanKane2010,QiZhang}. This type of behavior was theoretically predicted and experimentally observed for HgTe/CdTe quantum wells (2D-TI) and a few of narrow-gap semiconductors with strong spin-orbit interaction like Bi$_2$Se$_3$ and Bi$_2$Te$_3$ (3D-TI). The surface states in TIs are protected from backscattering in case of non-magnetic bulk or surface defects and have Dirac spectrum and so-called helical spin structure \cite{QiZhang}. 
 
 In Bi$_2$Se$_3$ the Dirac point is located inside the bulk energy gap, so having the chemical potential near the Dirac point is highly desirable not only to increase the ratio of surface to bulk conductivity, but more to work with small-wavevector states without any bulk impact. This opportunity can be crucial for potential realization of spintronic devices. However, attempts of stoichiometric growth of Bi$_2$Se$_3$ crystals have always given n-type of conductivity with the chemical potential located inside the bulk conduction band, see \cite{Huang2012}. For that reason the doping of the bulk is widely used to tune the chemical potential. It was shown, in partucular \cite{Beidenkopf2011}, that in the case of Mn or Ca doping, the bulk defects cause local fluctuations of the chemical potential so that its position varies substantially along the surface. In that case, one can't even define type of surface conductivity, so it might be a problem to create any electronic or spintronic devices using such a surface. 
 
To avoid this problem, we have grown Bi$_2$Se$_3$ crystal with an intrinsic conductivity type in a bulk without any doping, in the hope that the fewer number of bulk inhomogeneities will lead if not to the absence of the chemical potential fluctuations but at least to their much smaller value. Using the thermoelectric effect, we have found that our crystals have p-type of conduction. Scanning tunneling microscopy and spectroscopy (STM and STS) measurements demonstrate that measured with different $xy$--positioning of the tip, the chemical potential level of our crystals almost always sits inside the bulk energy gap within $\pm$50~meV from the Dirac point and is more often slightly shifted towards the valence band. As a result, a number of new features were found due to the absence of screening from the bulk. Namely, comparing the tunneling spectra obtained above clean surface without any visible defects and spectra measured above defects, one can see that they are undoubtedly different: the Dirac point remains at similar energies, but the density of filled and empty topological surface states changes. Also we have seen that the valence band shifts above the defect. This fact confirms that there is a local negative charge near the studied impurities. What is even more interesting, we have found that a 5ML-step at the surface of Bi$_2$Se$_3$ shifts the chemical level position near the step and change the shape of the tunneling spectra. The qualitative analysis of these variations is given. 
 
Bi$_2$Se$_3$ crystals were grown in an evacuated quartz ampoule from a stoichiometric mixture of Bi and Se with 3\% Se excess. The mixture was melted at $T=700^o$C, kept at this temperature for 10 hours and smoothly cooled down with the rate $-2^o$C/hour in the temperature gradient 1.5 K/cm along the ampoule. The resulting Bi$_2$Se$_3$ polycrystal was found to consist of approximately 0.1-1 mm sized monocrystalline blocks which were well suitable for STM study. The sign of the main carriers was found from the sign of the voltage appeared between the hot and cold copper blocks attached to a Bi$_2$Se$_3$ crystal: the potential of the cold block was larger,  in accordance with {\it p}-type of the current carriers. 

The Bi$_2$Se$_3$ crystalline structure consists of five-monolayer (5ML) slabs Se1-Bi1-Se2-Bi2-Se1' (quintuple layer) which can be easily splitted off from each other, see \cite{ZhangLiu2010}. The surface was prepared \textit{in situ} under ultra-high vacuum (UHV) conditions by cleavage at~$T=78$~K in the Omicron LT STM vacuum chamber with typical pressure $<5\cdot10^{-11}$~Torr. STM images were obtained in the constant current mode. For STS measurements we used tungsten tips with Au apex. 
The tip quality was controlled by measuring the contact resistance with a gold film  and and also a tunneling spectra of this film.
The obtained value of the resistance ($\lesssim 10^4{\rm\ \Omega}$) and linearity of the spectra confirm that the tip apex is metallic and appropriate for STS measurements. To obtain I-V curves for STS measurements we use 500 voltage points in each voltage sweep direction, no noticeable hysteresis between back and forth voltage sweeps was observed. The typical data set collected at any tunneling tip position consist of from 20 to 200 I-V curves which were averaged to reduce the noise. Differential conduction was calculated numerically from the resulting I–V curve.

\begin{figure}[h] 
 \includegraphics[width=.7\columnwidth]{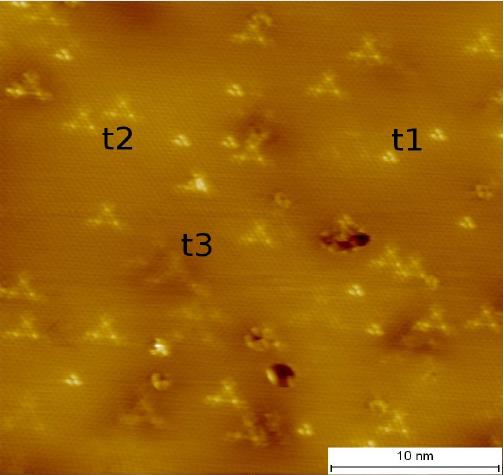} 
 \caption{STM image of Bi$_2$Se$_3$ surface with different types of defects. $V$=--0.4 V. All these types of defects vanish at positive voltages.}
 \label{defects}
 \end{figure}
 
Typical STM image of the sample surface is shown in Fig.~\ref{defects}. Three types of triangular defects of different size are discernible --- we call them \textbf{t1}, \textbf{t2} and \textbf{t3}. On the first glance the defects looks very similar to ones observed in n-type of Bi$_2$Se$_3$ \cite{Urazhdin2004}, but  the sizes and shapes of the defects are actually different. These defects are observed well at negative bias voltages, but completely disappear at positive ones. Furthermore, assuming that these triangular patterns are due to single substitutional atom located in subsurface layer which affects the local density of electronic states (LDOS) along $\sigma$-bonds \cite{Urazhdin2004}, one can conclude that these defects are in Bi1 and Bi2 layers, so they are likely to be Se substitutions. As far as we know such defects were not observed earlier. So, we attribute these features to subsurface Bi defects located in different Bi atomic layers, in contrast to Se defects reported earlier \cite{Urazhdin2004} and located in Se layers. Defects corresponding to the \textbf{t3}-type bigger triangles are located deeper beneath the surface. In general, these results of STM measurements are consistent with with the conclusion that the sample is of p-type.
 
 \begin{figure}[h] 
  \includegraphics[width=1\columnwidth]{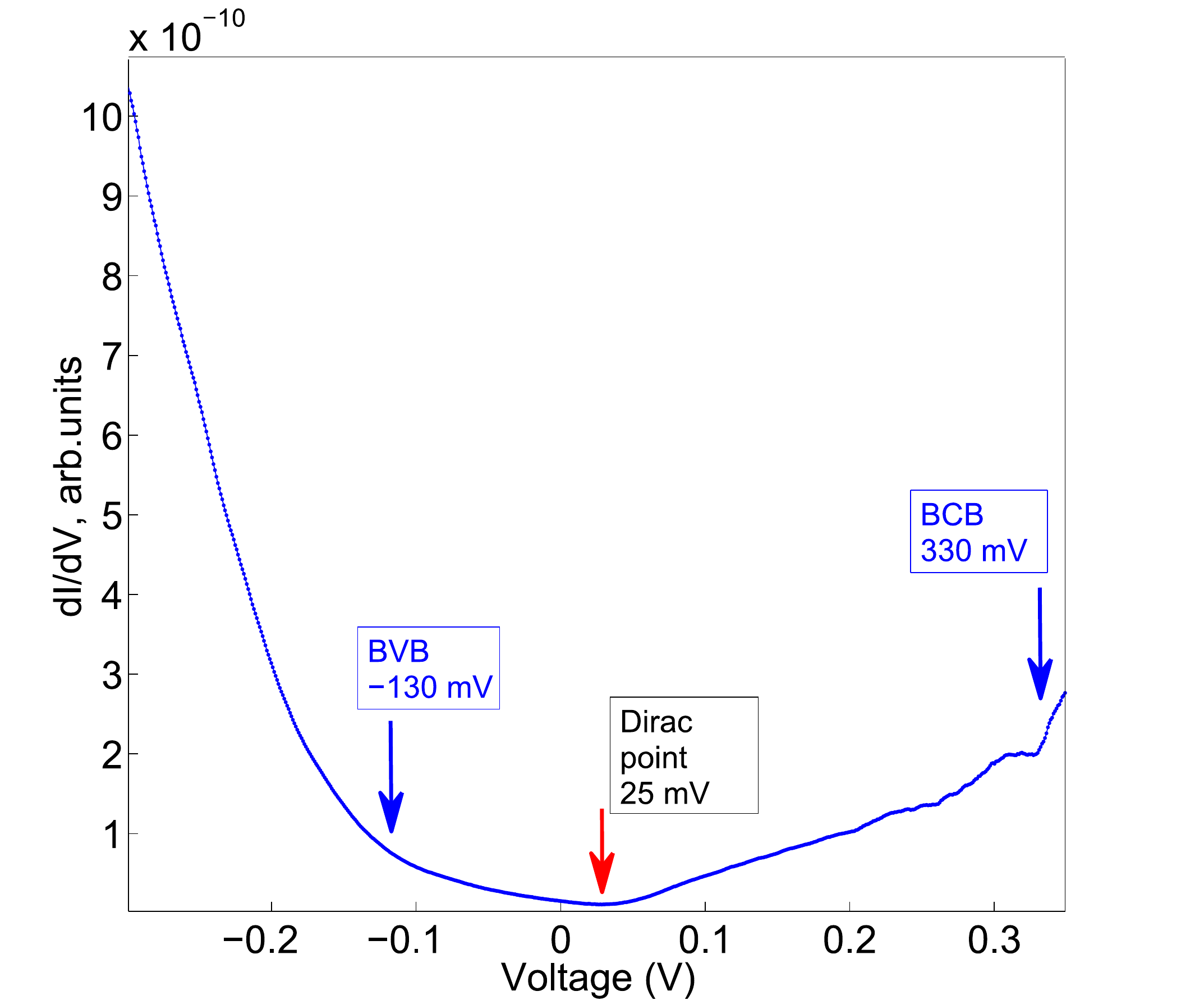}  
  \caption{STS-measurements above clean Bi$_2$Se$_3$ surface far from defects.}
  \label{cleanSTS}
 \end{figure}
 
The typical results of STS-measurements performed above the clean Bi$_2$Se$_3$ surface far from any defects are shown in Fig.~\ref{cleanSTS}. The spectra resembles one observed earlier \cite{Urazhdin2004} with the only difference: the chemical potential is located near the Dirac point and little closer to the bulk valence band (BVB). This also corresponds to the p-type conductivity, in agreement with the sign of thermoelectric effect and observed subsurface defects. We also found that when moving from point to point on a clean surface, the measured tunneling spectra are shifted by voltage, therefore the chemical potential is shifted too. The Dirac point position varies in a range from --100 to 25 mV from the chemical potential, so the latter is always located inside the bulk band gap (BBG). Thus studied Bi$_2$Se$_3$ crystal can be considered as an intrinsic semiconductor with slight predominance of p-type carriers.

\begin{figure}[h]  
\includegraphics[width=.7\columnwidth]{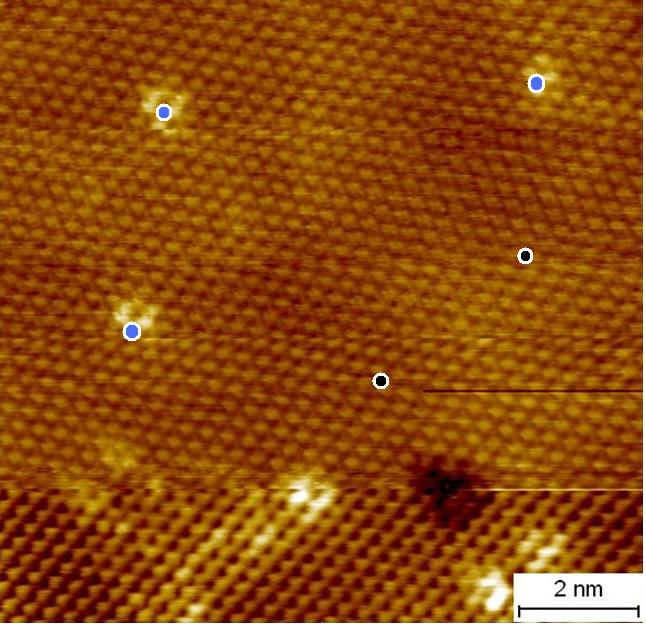}  
\caption{STM image ($I=100$~pA, $V=-400$~mV) of Bi$_2$Se$_3$ surface with \textbf{t1} defects. Blue/black points show places above/far from defects where tunneling spectra was measured.} \label{defs_pic}
\end{figure}

\begin{figure}[h]  
\includegraphics[width=1\columnwidth]{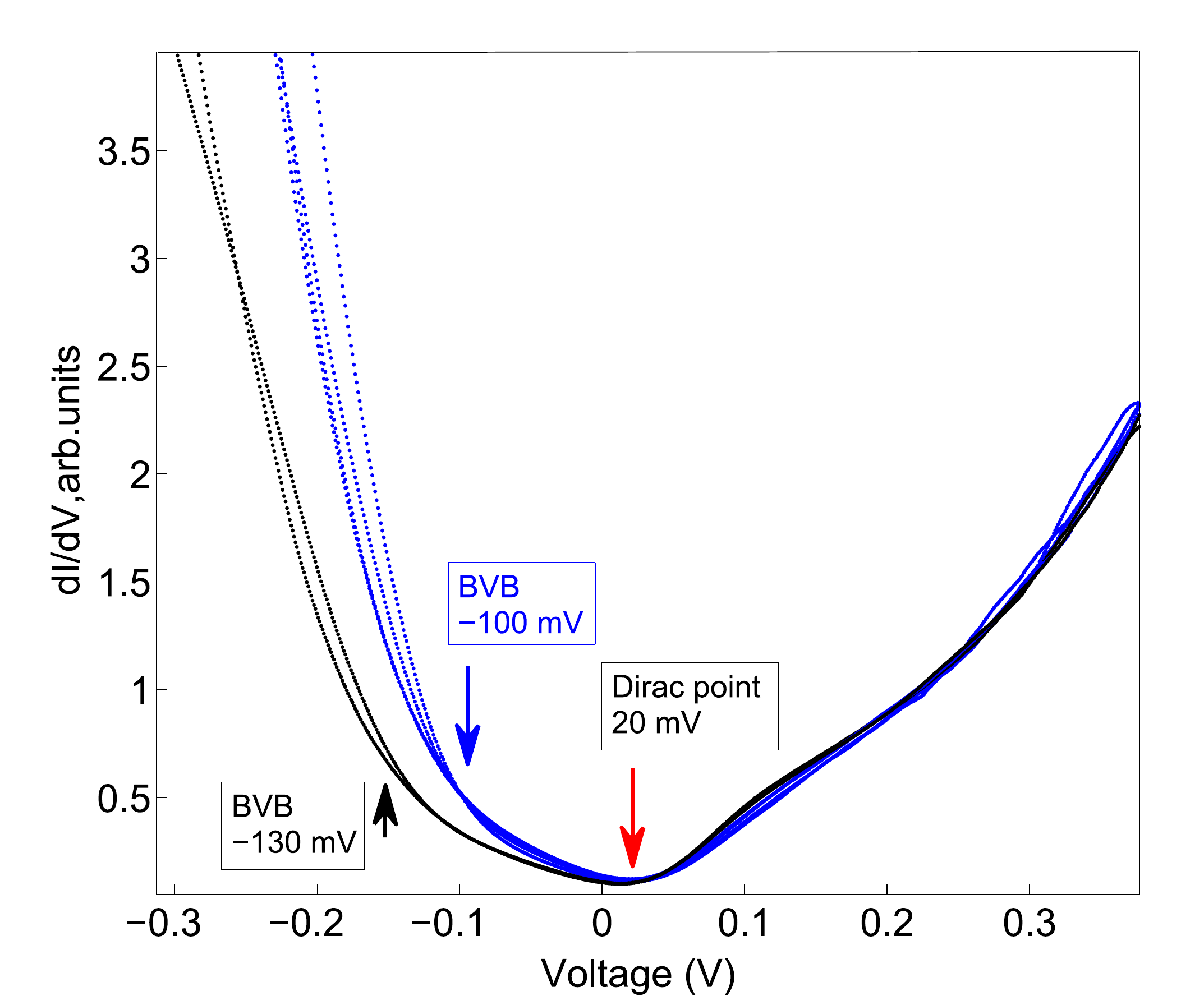}  
\caption{$dI/dU$ curves obtained above \textbf{t1} defects (blue) and far from them (black).} \label{defs}
\end{figure}

We also found a difference between the tunneling spectra taken above the \textbf{t1}-type defects and above the clean Bi$_2$Se$_3$ surface far from any defects. The obtained $dI/dV$ curves are shown in Fig.~\ref{defs}. We see that the BVB shifts to the negative voltages by approximately 30 mV. This can be explained by a local negative charge at substitutional atom, which is due to the acceptor type of this impurity. Although the Dirac point (DP) of the surface states spectrum remains almost at the same voltage, below DP ($V<V_D$) the local density of states (LDOS) above defects is higher, and over DP ($V>V_D$) the LDOS above defects is lower.
This illustrates the effect of subsurface defects on the surface LDOS.

\begin{figure}[h] 
\includegraphics[width=.7\columnwidth]{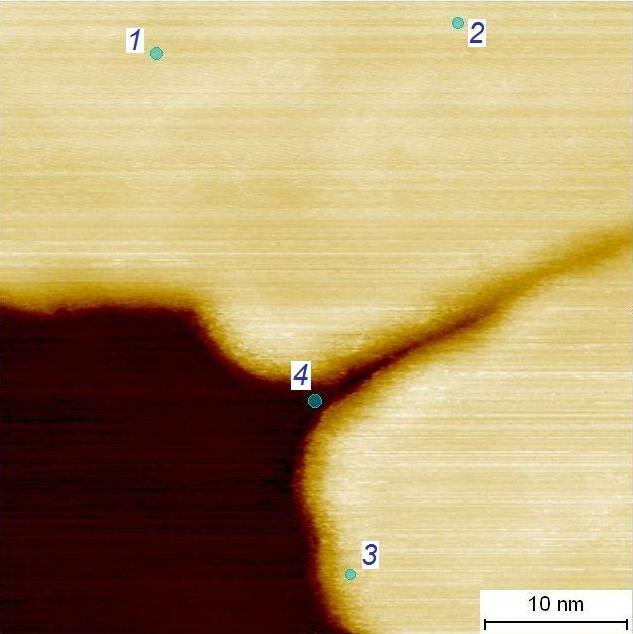}  
\caption{Quintuple step on the clean Bi$_2$Se$_3$ surface. Points where tunneling spectra were measured are marked by numbers.}
\label{StepY}
\end{figure}

\begin{figure}[h] 
\includegraphics[width=1\columnwidth]{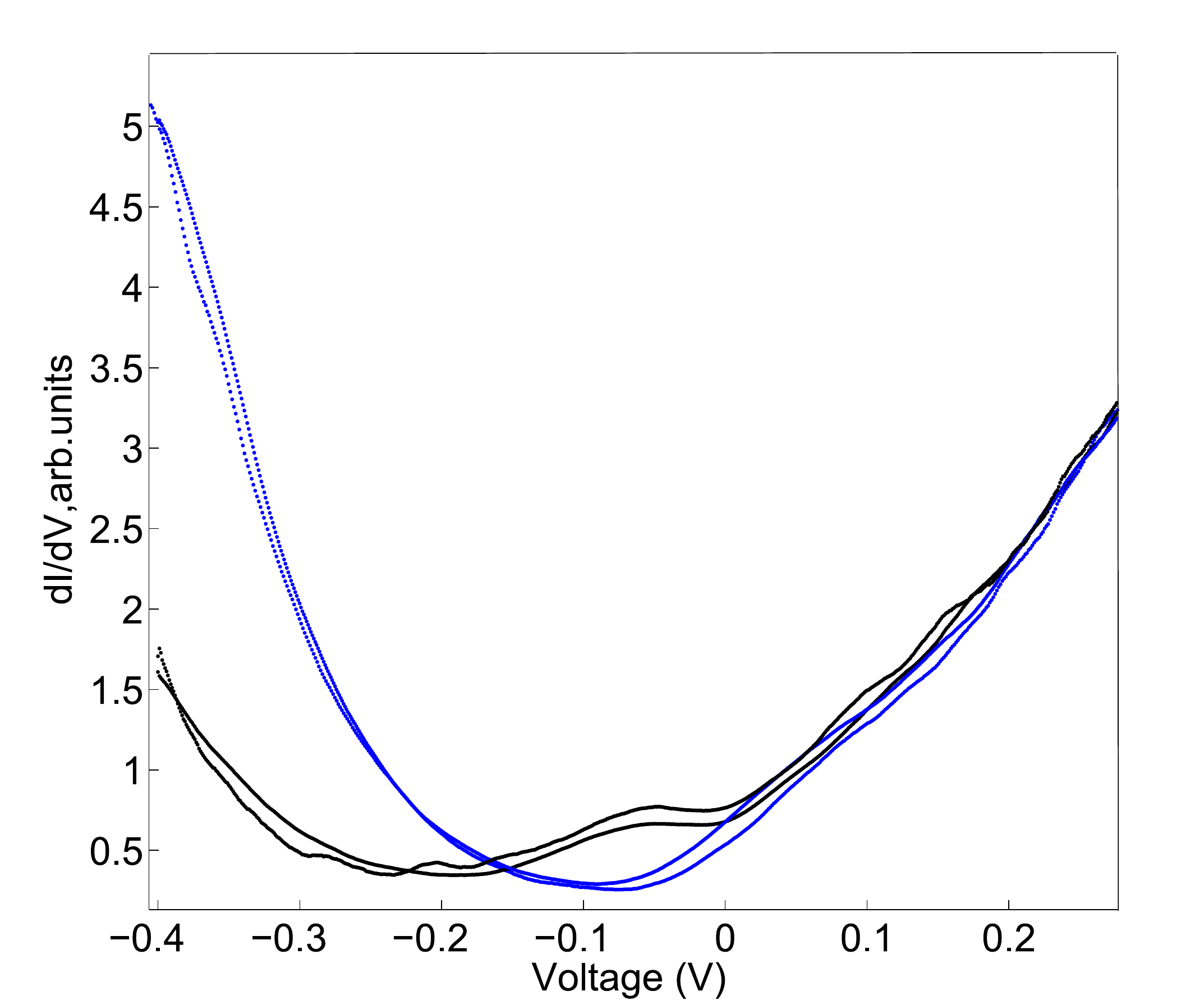} 
\caption{The $dI/dV$ curves measured at points  \# 1 and  \# 2 (see Fig.\ref{StepY}) far from the step edge (blue); at points  \# 3 and  \# 4 near the step edge (black).}
\label{StepVAH} 
\end{figure}

In addition, we have undertaken STS measurements near a step of height 0.8 nm (Fig.~\ref{StepY}) which corresponds to the height of the quintuple layer step. The tunneling spectra near the step edge was found to differ substantially from ones measured far from the step edge (see Fig.~\ref{StepVAH}). One can notice that the spectrum which was measured at the point \# 4 near the lower edge of the step coincides with the spectrum measured at the point \# 3 at the upper one, and spectra at the points \# 1 and \# 2 are also similar.As these two pairs of spectra are obviously different, so we can conclude that tunneling spectra change locally near the step edge. 

\begin{figure}[h] 
\includegraphics[width=0.7\columnwidth]{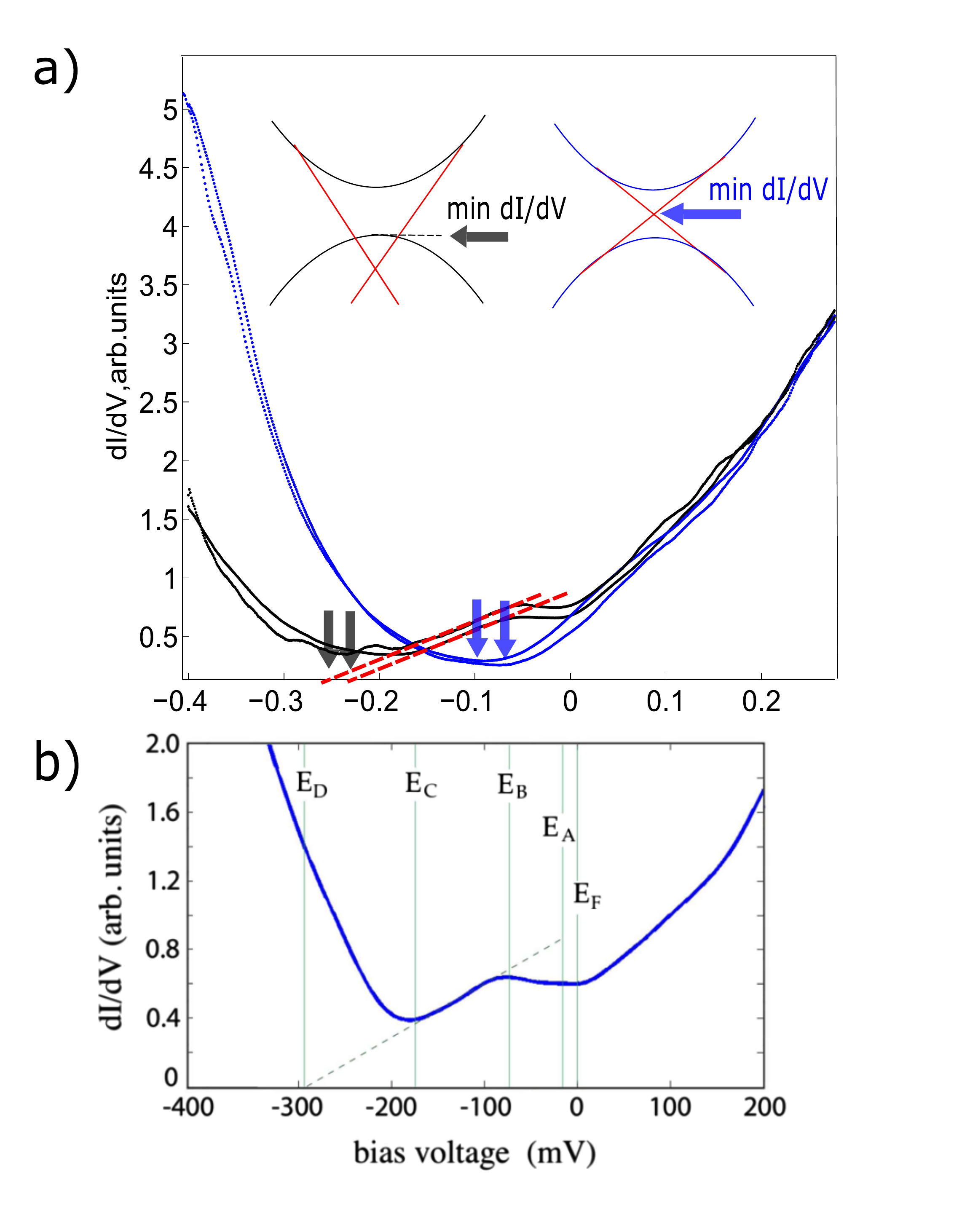}  
\caption{(a) $dI/dV$ curves obtained near (black ones) and far (blue ones) from the step. Black/blue arrows indicate the position of Dirac point of the surface states. The upper insets illustrate the relationship between the minimal value of tunneling conductance and position of Dirac point with respect to bulk band structure. (b) For comparison, Bi$_2$Te$_3$ surface spectra obtained in Ref.~\cite{Alpischev2010}. $E_C$ is the edge of BVB, $E_D$ is the energy of the Dirac point, $E_F$ is the chemical potential.}
\label{VAH_comp}
\end{figure}

To explain the physical sense of these changes, two approaches can be used. 
If we assume that the sharp rise of conductivity for the curves at Fig.~\ref{StepVAH} is due to the fact that the BVB or BCB states are involved into tunneling, then relatively large local change in the band gap width near the step edge must be admitted ($\Delta V_{gap}/V_{gap}\sim 20$\%), which does not look plausible. So we conclude that the BBG remains unchanged, but the Dirac cone of the surface states spectrum is shifted down by energy, and DP is located inside the BVB. It is clear that in this case the minimal value of $dI/dV$ corresponds to BVB, as it was demonstrated in \cite{Alpischev2010} comparing STM and ARPES measurements for Bi$_2$Te$_3$. Note that our spectrum measured near the edge is very similar to the one in \cite{Alpischev2010} (Fig.\ref{VAH_comp}). As the BBG for Bi$_2$Te$_3$ is~200~mV and the energy of the Dirac point is below BVB edge, so this coincidence looks surprising. 

\begin{figure}[h] 
\includegraphics[width=1\columnwidth]{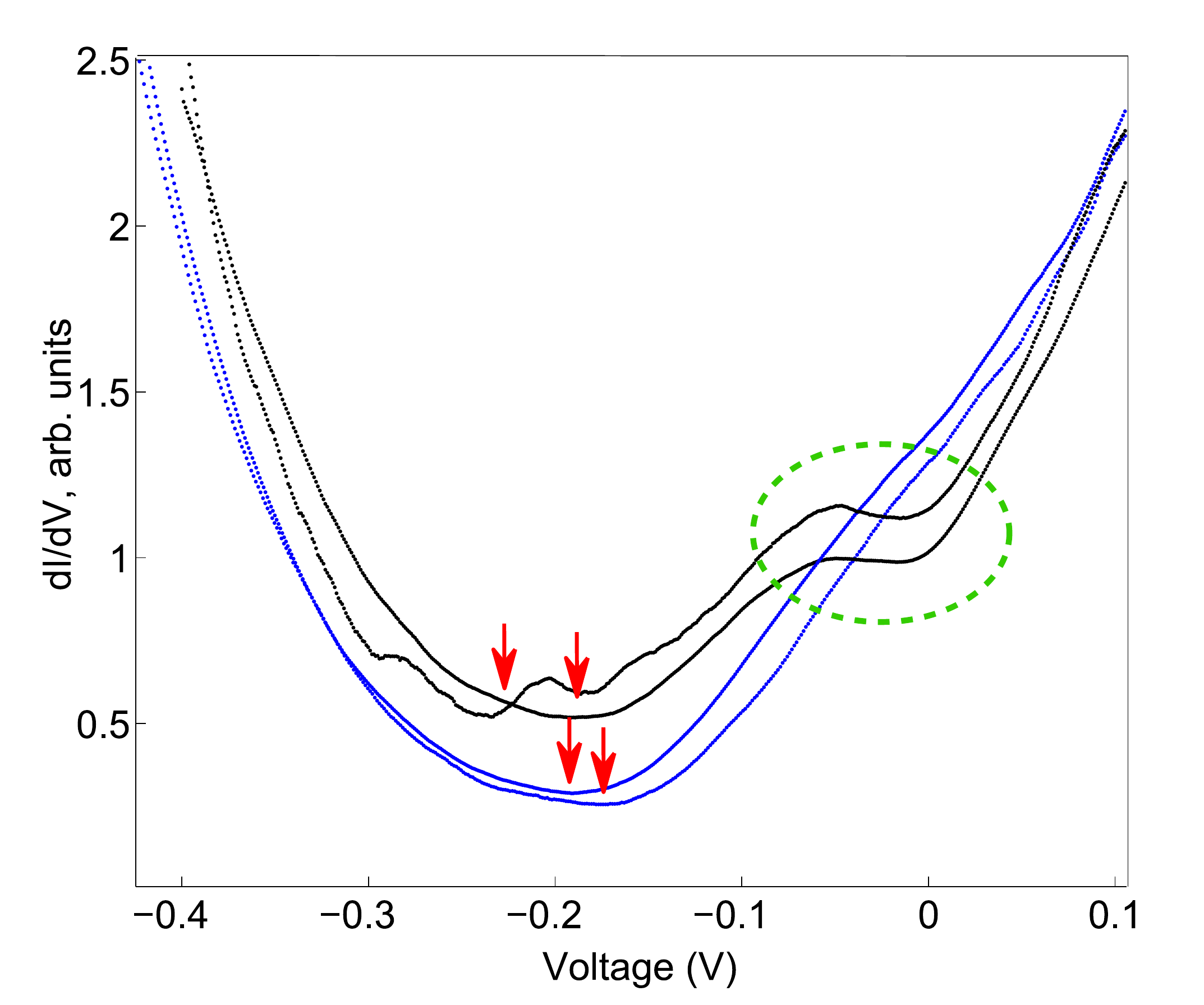} 
\caption{The same $dI/dV$ curves as in Fig.~\protect\ref{StepVAH} measured far from the step edge (at points\#  1 and \#  2 on Fig.\ref{StepY}) (blue); near the step edge (at points \# 3 and \# 4) (black). The curves are normalized in a way different from the curves at Fig. \ref{StepVAH} (see text for details). Spectra measured far from the edge are shifted horizontally by subtracting 100~mV. Red arrows indicate positions of the Dirac points. Green oval mark the differences between the spectra.}
\label{StepVAH_chemicalshift} 
\end{figure}

Alternatively, the I-V curves obtained in STS measurements can be normalized in a way that gives the same BBG value. The curves at Fig. \ref{StepVAH_chemicalshift} are normalized to make BBG widths equal for all the spectra, the edges of BBG are supposed to be at the voltages since which the intensive and steady raise of conductivity is noticed, namely, at voltage band edges.
The curves at Fig.~\ref{StepVAH_chemicalshift} measured far from the edge are shifted here by subtracting 100~mV. We can see the alignment of Dirac points for spectra far and close to the step edge, which means that the step edge results in the shift of the chemical potential by -100 mV. Anyway, the origin of such changes in the tunneling spectra and BBG near the step edge could be clarified by \textit{ab initio} calculations which are beyond the scope of this work.

Summarizing the results, UHV low-temperature STM study of atomically clean p-type Bi$_2$Se$_3$(111) surface prepared \textit{in situ} have shown the presence of subsurface defects in Bi layers. 
STS-measurements demonstrate that the chemical potential is always located inside the bulk energy gap, but slightly varies along the surface. 
It was found that the observed subsurface defects as well as the quintuple layer step edge causes the significant changes in STS spectra. Comparing tunneling spectra obtained above the clean surface without any visible defects and spectra measured above defects, one can see that they are undoubtedly different: the Dirac point remains at similar energies, but the density of filled and empty topological surface states changes. The valence band shift was also observed above the defects. This fact confirms that there is a local negative charge near the studied impurities. In addition, we observed the substantial variations in tunneling spectra near the quintuple edge accompanied whether by the shift of the Dirac point with respect to the bulk energy gap of Bi$_2$Se$_3$, or by the shift of the chemical potential with respect to the Dirac point. 

\begin{acknowledgments}
The work was supported by the Russian Foundation for Basic Research  and the program of Physical Department of RAS. 
\end{acknowledgments}

 \bibliography{HasanKane2010,QiZhang,Huang2012,Beidenkopf2011,Alpischev2010,Urazhdin2004,ZhangLiu2010}
\newpage

\newpage

\newpage

\newpage

\newpage

\newpage

\end{document}